\documentclass[prd,preprint]{revtex4}
\usepackage[utf8]{inputenc}
\usepackage{latexsym}
\usepackage{epsfig}
\usepackage{graphicx}
\usepackage[normalem]{ulem}
\usepackage{color}
\usepackage{xcolor}
\usepackage{multirow}
\usepackage{hyperref}
\usepackage{numprint}
\usepackage{eucal}
\usepackage{amsmath}
\usepackage{amssymb}

	\usepackage{makeidx}
	\usepackage{amsfonts}
	\usepackage[ansinew]{}
	\usepackage[usenames,dvipsnames]{pstricks}
	\usepackage{epsfig}
	\usepackage{float}
\usepackage[english]{babel} 
\usepackage{subcaption}
 \renewcommand\theequation{\arabic{equation}}
\newcommand{\be}{\begin{equation}}
\newcommand{\ee}{\end{equation}}
\newcommand{\bea}{\begin{eqnarray}}
\newcommand{\eea}{\end{eqnarray}}

\usepackage{morefloats} 
\usepackage[titletoc,title]{appendix}
\usepackage{color}

\begin{document}
\title{Effective speed approach for modified scalar field propagation}
\author{Kevin Restrepo Tobón}
\author{Antonio Enea Romano}
\email{Corresponding author: Antonio Enea Romano, antonioenea.romano@ligo.org}
\affiliation{Instituto de Fisica, Universidad de Antioquia, A.A.1226, Medellin, Colombia}
\begin{abstract}
    We study the propagation of a constant-speed Gaussian scalar field wave packet (GWP) in Minkowski space, showing that the energy conditions are violated for superluminal speeds. 
    We then apply the effective speed approach to the GWP propagation, deriving the corresponding effective metric, effective Lagrangian and effective stress-energy tensor, and test the energy conditions defined in terms of the effective metric.
    We study the nature of the effective source driving  the speed modification, and find that it corresponds to a perfect fluid with negative equation of state.
\end{abstract}
\maketitle

\section{Introduction}
Cosmological perturbations play a fundamental role in understanding the evolution of the Universe, since they allow us to model inhomogeneities from which the large scale structure has formed.
It has been shown \cite{romano_effective_2024} how it is possible to encode the effects of the interaction of cosmological perturbations with other fields in an effective speed, but this effective approach can also be applied to other fields satisfying a similar equation of motion. In this paper we apply the effective speed approach to a scalar field propagating in Minkowski space, specifically to a Gaussian wave packet propagating at constant speed.
We compute the effective speed and the effective metric, showing that this effective description provides equations that admit the same solutions as the equations with a source term.
We then test several energy conditions, comparing the results obtained with the Minkowski metric and the effective metric.

\section{Effective source of a Gaussian wave packet}
The Minkowski metric in spherical coordinates takes the form
\begin{align}
    \eta_{\mu\nu} = 
    \begin{pmatrix}
        c^2 & 0 \\
        0 & -\Omega_{ij}
    \end{pmatrix}\,;\qquad
    \Omega_{ij} = \begin{pmatrix}1 & 0&0\\0&r^2&0\\0&0&r^2\sin^2\theta\end{pmatrix}\,. \label{eq:minkowski_metric}
\end{align}
For a scalar field with Lagrangian
\begin{equation}
\mathcal{L} =\frac{1}{2}\partial_\mu\phi\partial^\mu\phi =\frac{1}{2}\left[\frac{\dot\phi^2}{c^2}-(\nabla \phi)^2\right]\,,
\label{eq:free-lagrangian}
\end{equation}
the equation of motion is given by the d'Alembert operator
\begin{equation}
\Box \phi = \frac{1}{c^2}\ddot\phi-\nabla^2\phi=0\,,
\end{equation}
where the dot represents a time derivative, $\nabla^2 = \partial_i\partial^i$ is the Laplacian and $c$ is the speed of light. 
Given a Gaussian wave packet of the form
\be
\phi_g(t,r) =\frac{\phi_0}{r} \exp{\left[-\frac{(r-c_gt)^2}{\sigma^2}\right]}\,,\label{eq:GWP}
\ee
we define the function $\Pi_g(t,\mathbf{r})$ as 
\begin{gather}
    \Pi_g(t,r) = c^2\,\Box\phi_g = \frac{2\phi_g(t,r)}{\sigma^4}\left(c_g^2-c^2\right) \left[2(r-c_gt)^2-\sigma ^2\right]\,. \label{eq:PI}
\end{gather}
By construction $\phi_g$ is a solution of the inhomogeneous equation
\begin{equation}
\ddot\phi_g-c^2\nabla^2\phi_g = \Pi_g\,, \label{eq:EOM}
\end{equation}
which can be interpreted as the source term necessary to obtain a Gaussian wave packet propagating in Minkowski space at a constant speed $c_g$.
The corresponding Lagrangian density is
\begin{equation}
\mathcal{L} = \frac{1}{2}\left[\frac{\dot\phi^2}{c^2}-(\nabla \phi)^2+2\frac{\Pi_g \phi}{c^2}\right]\,.\label{eq:minkowski-lagrangian}
\end{equation}



\section{Energy conditions computed with the Minkowski metric}

In this section we study several energy conditions \cite{kontou_energy_2020,hawkingSingularitiesGravitationalCollapse1970} for the GWP. In the following  we denote $\phi_g$ as $\phi$ for ease of notation, assuming an arbitrary propagation speed $c_g$.
In order to test the energy conditions we need the stress-energy tensor (SET) \cite{misnerGravitation2008} for the Lagrangian in Eq.~(\ref{eq:minkowski-lagrangian}), which is given by
\begin{align}
T_{\mu\nu} &~:=~ \frac{2}{\sqrt{|\eta|}}\frac{\delta S}{\delta \eta^{\mu\nu}}=\partial_\mu \phi\partial_\nu \phi -\eta_{\mu\nu}\mathcal{L} \label{eq:stress-energy-tensor}\\
&= \partial_\mu \phi\partial_\nu \phi - \frac{1}{2}\eta_{\mu\nu}
    \left[\frac{\dot{\phi}^2}{c^2}-(\nabla\phi)^2+\frac{2\Pi_g\phi}{c^2}\right].\nonumber
\end{align}

\subsection{Null energy condition}
The null energy condition (NEC) states that for any null vector $k^\mu$, defined by $k^\mu k_\mu = 0$
\begin{align}
    T_{\mu\nu}k^\mu k^\nu \geq 0\,.
\end{align}
From the above equations we have
\begin{align}
    T_{\mu\nu}k^\mu k^\nu &= \left(\partial_\mu \phi\partial_\nu \phi -\eta_{\mu\nu}\mathcal{L}\right)k^\mu k^\nu\\
    &=\left(\partial_\mu\phi k^\mu\right)^2\nonumber\,,
\end{align}  
which is positive definite \cite{barceloScalarFieldsEnergy2000,visser_energy_2000}, implying that the NEC is always satisfied.

\subsection{Weak energy condition}
For any timelike vector $t^\mu$, defined by $t^2=\eta_{\mu\nu}t^\mu t^\nu=t^\mu t_\mu >0$,
the weak energy condition (WEC) states that
\begin{align}
    T_{\mu\nu}t^\mu t^\nu \geq 0\,. \label{eq:WEC}
\end{align}


A general timelike vector can be parameterized in the following way
\begin{equation}
    t^\mu = \frac{1}{\sqrt{1-\beta^2}}(1/c,\beta\, n^i)\,,
\end{equation}
where $\lvert\beta\rvert<1$, $n^i$ satisfies $n^in_i=1$, and by construction $t^\mu t_\mu = 1$. 
Replacing $T_{\mu\nu}$ in Eq.~(\ref{eq:WEC}) we get
\begin{equation}
   T_{\mu\nu}t^\mu t^\nu  =  (\partial_\mu\phi t^\mu)^2 - \frac{1}{2}\left[\frac{\dot\phi^2}{c^2}-(\nabla\phi)^2+\frac{2\Pi_g\phi}{c^2}\right]\,\label{eq:WEC-minkowski}.
\end{equation}

By substituting $\Pi_g$, $\phi_g$, and $t^\mu$ we can use a combination of algebra and calculus to show that for $c_g>c$ one can find at least one observer for which Eq.~(\ref{eq:WEC-minkowski}) is negative for some region of space and time, thus violating the WEC, as shown in more detail in appendix \ref{ap:WEC-minkowski}.

\subsection{Strong energy condition}

The strong energy condition (SEC) states that for any timelike vector $t^\mu$ 
\begin{equation}
    (T_{\mu\nu}-\frac{1}{2}T \eta_{\mu\nu} )t^\mu t^\nu\geq 0\,.\label{eq:SEC}
\end{equation}
From Eq.~(\ref{eq:stress-energy-tensor}) we get
\begin{equation}
    (T_{\mu\nu}-\frac{1}{2}T \eta_{\mu\nu} )t^\mu t^\nu=(\partial_\mu\phi t^\mu)^2+\frac{\Pi_g\phi}{c^2}\,,\label{eq:SEC-part2}
\end{equation}
Performing an analysis similar to that in appendix \ref{ap:WEC-effective}, one finds that the SEC is violated locally inside the GWP when $c_g>c$. 


\section{\label{sec:source_tensor}Driving engine of the speed modification}
The SET of the source term can be computed by subtracting  from the total one defined in Eq.~(\ref{eq:stress-energy-tensor}) that corresponding to a free scalar field
\begin{align}
    T^{\mathrm{free}}_{\mu\nu} &= \partial_\mu\phi\partial_\nu\phi-\eta_{\mu\nu}\mathcal{L}_{\mathrm{free}}= \partial_\mu\phi\partial_\nu\phi-\frac{1}{2}\eta_{\mu\nu}\left[\frac{\dot\phi^2}{c^2}-(\nabla\phi)^2\right]\,,\nonumber
\end{align}
obtaining
\begin{equation}
T^{\mathrm{source}}_{\mu\nu} = T_{\mu\nu} -T^{\mathrm{free}}_{\mu\nu} = \rho_s c^2 \eta_{\mu\nu}\,,\label{eq:src_state}
\end{equation}
where 
\begin{equation}
    \rho_s = -\frac{\Pi_g \phi}{c^4}.
\end{equation}
The  SET of a spherically symmetric perfect fluid is given by
\be
T_{\mu\nu}=\Big[\rho(r)+\frac{P(r)}{c^2}\Big]U_{\mu}U_{\nu} - P(r) \eta_{\mu\nu}\,,
\ee
which compared to Eq.~(\ref{eq:src_state}) 
shows that the SET of the source term, which is the driving engine modifying the speed of the wave packet, corresponds to a perfect fluid with  equation of state $P_s=-\rho_s c^2$, i.e. it can be interpreted as a spherically symmetric dark energy pulse. 

\begin{figure}[h!]
    \includegraphics[width=\linewidth]{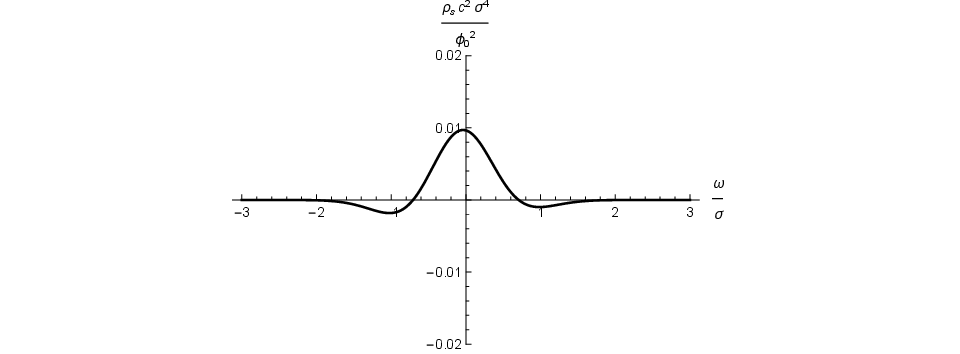}
 \caption{Plot of the source term energy density $\rho_s$ for $c_g = 1.1 c$. We have defined $\omega = (r-c_g t)$.}
 \label{fig:rho}
\end{figure}

\begin{figure}[h!]
    \includegraphics[width=\linewidth]{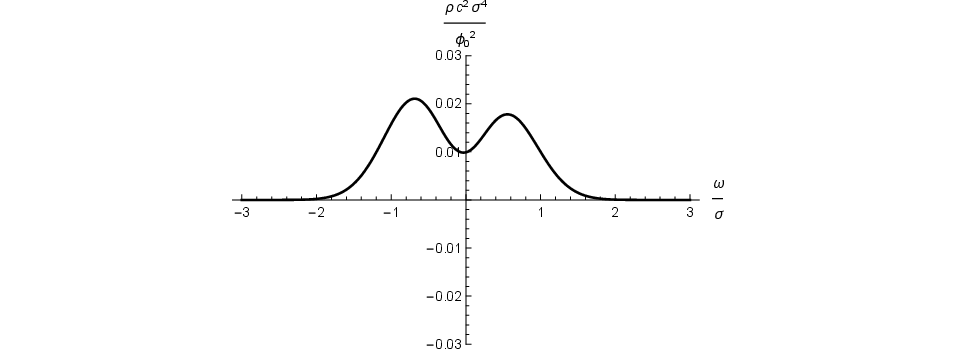}
 \caption{Plot of the total energy density $\rho$ corresponding to Eq.(\ref{eq:stress-energy-tensor}) for a static observer, and for $c_g = 1.1 c$. We have defined $\omega = (r-c_g t)$.}
 \label{fig:rho_total}
\end{figure}

In Fig.~(\ref{fig:rho}) we show the energy density profile of the driving engine, as a function of the radial coordinate, around the center of the wave packet.   Note that  while the driving engine has some regions with negative $\rho_s$, the total energy density is always positive, as shown in Fig.~(\ref{fig:rho_total})

\section{\label{sec:effective_approach}Effective speed approach}
We can study the solution of Eq.~(\ref{eq:EOM}) using an effective approach as done in \cite{romano_effective_2024,ROMANO2024101492}, where the effects of the source term are encoded in a space and time-dependent effective speed $c_e$. In the following we briefly review the definition of the effective speed, and then we apply this approach to the GWP.

Let us consider an equation of the form
\be
\ddot{\phi}-c^2\nabla^2\phi=\Pi(\phi) 
\ee
denote a solution by $\hat\phi$, and define $\hat\Pi=\Pi(\hat\phi)$.
We can manipulate the equation as follows
\begin{equation}
    \ddot{\hat\phi}-c^2\nabla^2\hat\phi-\hat\Pi = \partial_t\left[\dot{\hat\phi}-\int \hat\Pi \mathrm{d}t\right]-c^2\nabla^2\hat\phi =\partial_t\left[\dot{\hat\phi}\left(1-\frac{\hat g}{\dot{\hat\phi}}\right)\right]-c^2\nabla^2\hat\phi = 0\,,
\end{equation}
where we have defined $\hat g = \int \hat\Pi\, \mathrm{d}t$. After defining the effective speed as
\begin{equation}
    c_e^2 = c^2\left(1-\frac{\hat g}{\dot{\hat\phi}}\right)^{-1}\,,\label{eq:ce}
\end{equation}
the equation can be rewritten as
\begin{equation}
\partial_t\left(\dot{\hat\phi}\frac{c^2}{c_e^2}\right)-c^2\nabla^2\hat\phi=0\,,
\end{equation}
from which we finally get
\be
\ddot{\hat\phi}-2\frac{\dot c_e}{c_e}\dot{\hat\phi}-c_e^2\nabla^2\hat\phi = 0\,.\label{eq:eom_effective}
\ee
A solution of Eq.~(\ref{eq:EOM}) is by construction also a solution of Eq.~(\ref{eq:eom_effective}). 
    
\subsection{Effective metric}
It is possible to define an effective metric such that the Lagrangian of a free field gives the effective equation. This metric takes the form
\begin{align}
    g^{\mathrm{eff}}_{\mu\nu} = 
    \begin{pmatrix}
        c_e & 0 \\
        0 & -\frac{1}{c_e}\Omega_{ij}
    \end{pmatrix}\,.
\end{align}


It is easy to check that for the free field Lagrangian 
\begin{equation}
\mathcal{L}_{\mathrm{eff}} = \frac{1}{2}g_{\mathrm{eff}}^{\mu\nu}\partial_\mu\phi\partial_\nu\phi\,,
\end{equation}
the Lagrange equations give the same effective equation given in Eq.~(\ref{eq:eom_effective})
\begin{align}
    \Box \phi = \frac{1}{\sqrt{-g_\mathrm{eff}}}\partial_\mu(\sqrt{-g_\mathrm{eff}}\partial^\mu\phi) = \ddot\phi-2\frac{\dot{c_e}}{c_e}\dot\phi-c_e^2\nabla^2\phi = 0 \label{eq:EOM_effective}.
\end{align}

\subsection{Effective energy-momentum tensor}
From variation of the effective action with respect to the effective metric we obtain the effective energy-momentum tensor

\begin{align}
T^{\mathrm{eff}}_{\mu\nu} &= \frac{2}{\sqrt{|g_\mathrm{eff}|}}\frac{\delta S_\mathrm{eff}}{\delta g_\mathrm{eff}^{\mu\nu}}= \partial_\mu \phi\partial_\nu \phi -g^{\mathrm{eff}}_{\mu\nu}\mathcal{L}_{\mathrm{eff}}\label{eq:stress-energy-tensor-effective} \\
&= \partial_\mu \phi\partial_\nu \phi - 
\frac{1}{2}
    \begin{pmatrix}
        c_e & 0 \\
        0 & -\frac{1}{c_e}\Omega_{ij}
    \end{pmatrix}
    \left[\frac{\dot{\phi}^2}{c_e}-c_e(\nabla\phi)^2\right].\nonumber
\end{align}
\section{Effective approach for the Gaussian wave packet}
Substituting $\phi_g$ and $\Pi_g$ into the definition of $c_e$ gives
\begin{align}
    c_e^2 = c^2\left(1-\frac{\int \Pi_g \mathrm{d}t}{\dot{\phi_g}}\right)^{-1} 
    &= c^2\left[1-\frac{2\phi_g(c_g^2-c^2)(r-c_gt)/(c_g\sigma^2)}{2\phi_gc_g(r-c_gt)/\sigma^2}\right]^{-1} \\
    &= c^2\left(1-\frac{c_g^2-c^2}{c_g^2}\right)^{-1}=c^2\left(\frac{c^2}{c_g^2}\right)^{-1}
    = c_g^2\,,\nonumber
\end{align}

In this case the effective speed is the quantity $c_g$ appearing in the GWP, i.e. it is the speed of the wave peak, confirming the physical meaning of the effective speed.
Note that in this case the friction term is absent in the effective equation, because the effective speed is constant.
This result is expected, since $\phi_g$ is a solution of the effective equation 
\be
\ddot{\phi}-c_g^2\nabla^2\phi=0\,, 
\ee
which can be obtained directly from the d'Alembert operator defined in terms of the effective metric. 
In the effective speed description there is no source term in the wave equation, but the metric from which the free Lagrangian is defined is not the Minkowski metric, but the effective metric. 


\subsection{Energy conditions for the effective stress-energy tensor}
\subsubsection{NEC}
Defining the null vector $k^\mu$ in terms of the effective metric, i.e. assuming $g_{\mu\nu}^{\mathrm{eff}}k^\mu k^\nu = 0$, we obtain 
\be
T^{\mathrm{eff}}_{\mu\nu}k^\mu k^\nu= (\partial_\mu\phi k^\mu)^2\,,
\ee
which is always positive, implying the NEC is always satisfied.

\subsubsection{WEC}
Defining $t^\mu$ as a timelike vector in terms of the effective metric, i.e. assuming $t^2 = g^{\mathrm{eff}}_{\mu\nu}t^\mu t^\nu>0$, the WEC gives
\begin{align}
    T^{\mathrm{eff}}_{\mu\nu}t^\mu t^\nu &= \left(\partial_\mu \phi\partial_\nu \phi -g^{\mathrm{eff}}_{\mu\nu}\mathcal{L}_{\mathrm{eff}}\right)t^\mu t^\nu\\
    &=\left(\partial_\mu\phi t^\mu\right)^2- t^2\mathcal{L}_{\mathrm{eff}} \geq0 \nonumber.
\end{align}

A general unit timelike vector, defined by the condition $g^{\mathrm{eff}}_{\mu\nu}t^\mu t^\nu = 1$, can be parameterized as
\begin{equation}
t^\mu = \frac{\sqrt{c_e}}{\sqrt{1-\beta^2}}(\frac{1}{c_e},\beta n^i)\,,
\end{equation}
where $|\beta|<1$ and $n^i = (n_r,n_\theta,n_\phi)$ are the components of a vector satisfying $n^i n_i=1$. In appendix \ref{ap:WEC-effective} it is shown that the WEC is satisfied for any vector $n^i$ and any value of $\beta$.
\subsubsection{SEC}
Assuming a timelike vector in terms of the effective metric, i.e. satisfying the condition $t^2 = g^{\mathrm{eff}}_{\mu\nu}t^\mu t^\nu=1$, the SEC takes the form
\begin{equation}
    (T^{\mathrm{eff}}_{\mu\nu}-\frac{1}{2}T^{\mathrm{eff}} g^{\mathrm{eff}}_{\mu\nu} )t^\mu t^\nu=(\partial_\mu\phi t^\mu)^2\,,
\end{equation}
which is always positive, implying the SEC is always satisfied.

\section{Conclusions}
We have studied the energy conditions for a Gaussian wave packet propagating at constant speed in Minkowski space, confirming \cite{visserSuperluminalCensorship2000,curielPrimerEnergyConditions2017,barceloTwilightEnergyConditions2002,santiagoGenericWarpDrives2022,barceloTraversableWormholesMassless1999,flanaganDoesBackReaction1996} that superluminal speeds lead to a violation of the energy conditions. We have assumed the field satisfies a wave equation with the d'Alembert operator determined by the Minkowski metric, and computed the source term that gives wave packet propagation at an arbitrary constant speed. Once the source term has been computed, we calculated the corresponding Lagrangian, and from it the stress-energy tensor, assuming the Minkowski metric.

We then formulated an equivalent mathematical description of the system, based on the effective speed and effective metric.
This effective description gives equations of motion that admit the same solutions, but without source terms. The effective metric allows  to obtain an effective Lagrangian that gives the effective equation of motion without a source term. 
We then computed the effective stress-energy tensor from the effective Lagrangian and metric, and studied the energy conditions in terms of the effective metric.
In order understand what is the nature of the driving engine which modifies the wave packet propagation we have studied the stress energy tensor of the source term, finding that it can be interpreted as perfect fluid with negative equations of state. In the future it will be interesting to consider the propagation on an expanding universe cosmological background.

\section{Author Contributions}
A.E.R. and K.R.T. contributed to writing the manuscript and to the calculations. K.R.T. created the figures.


\section{Data Availability}
All data generated or analyzed during this study are included in this published article or are available from the corresponding author on reasonable request.

\newpage
\appendix
\counterwithin*{equation}{section}
\renewcommand\theequation{\thesection\arabic{equation}}

\section{\label{ap:WEC-minkowski}WEC with the Minkowski metric}
\begin{figure}[h]
    \includegraphics[width=\linewidth]{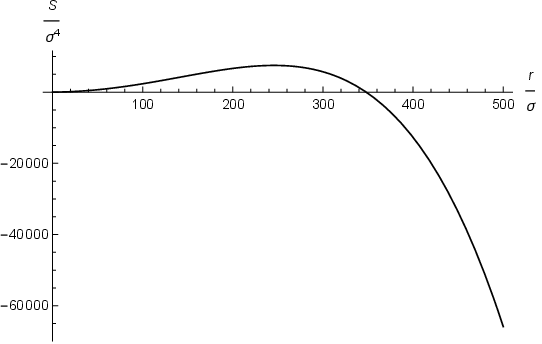}
 \caption{The quantity $S$ is plotted as a function of $r$, for  $\alpha = 1.1$,  $\beta = \beta_m+10^{-6}$, $t = 10 \sigma/c$ and $n_r = -1$.}
 \label{fig:minkowski-wec}
\end{figure}

Substituting $t^\mu = \frac{1}{\sqrt{1-\beta^2}}(1/c,\beta n^i)$ and Eq.~(\ref{eq:stress-energy-tensor}) in the WEC gives
\begin{align}
    T_{\mu\nu}t^\mu t^\nu = (\partial_\mu\phi_g t^\mu)^2-\mathcal{L}
    =\frac{\phi_g^2}{r^2\sigma^4}\left(\frac{E_0}{1-\beta^2}-\frac{E_1}{2}\right)\,,\label{eq:ap-WEC-general}
\end{align}
where we used $\eta_{\mu\nu}t^\mu t^\nu = 1$, defined $c_g = \alpha \, c$, amd $E_0$ and $E_1$ are given by
\begin{align}
E_0 &=  \left[n_r \beta  \sigma ^2+2 r (n_r \beta +\alpha ) (r-\alpha  c t)\right]^2\,,\\
E_1 &= 12r^2(r-\alpha c t )^2(\alpha^2-1)-\sigma^2\left[4\alpha^2r(r-\frac{ct}{\alpha})+\sigma^2\right]\,.
\end{align}
The sign of Eq.~(\ref{eq:stress-energy-tensor}) is determined by
\begin{equation}
     S = \frac{E_0}{1-\beta^2}-\frac{E_1}{2}\,.
\end{equation}
We now proceed to show that there are observers for which  $S<0$ for any $\alpha>1$.
Let's set $n_r = -1$  and consider the leading terms in the limit $(r-c t \alpha)\to\infty$ 
\begin{align}
    S  &\approx 2r^2(r-\alpha c t)^2\left[\frac{2(\alpha-\beta)^2}{1-\beta^2}-3(\alpha^2-1)\right]\,.
\end{align}
The sign of the above expression is determined by the last term, which is quadratic in $\beta$. We can find the values of $\beta$ for which $S<0$ by solving the equation
\begin{equation}
    \frac{2(\alpha-\beta)^2}{1-\beta^2}-3(\alpha^2-1)=0\,,
\end{equation}
and consider the solution 
\begin{equation}
    \beta_m = \frac{2 \alpha - \sqrt{3\left(\alpha ^4-2 \alpha ^2+1\right)}}{3 \alpha ^2-1}\,.
\end{equation}
For  $\alpha>1$,  and $\beta>\beta_m$  the WEC is violated for any large enough $r$, as shown in Fig.~(\ref{fig:minkowski-wec}).


\section{\label{ap:WEC-effective}WEC with effective metric}
The sign of $T^{\mathrm{eff}}_{\mu\nu}t^\mu t^\nu $ is given by the sign of the expression
\begin{equation}
    S=8 r^2 (\beta  n_r-1)^2 (r-\alpha  c t)^2+4 r \sigma ^2 \left[2 n_r \beta\left(\beta  n_r-1\right)-\beta^2 +1\right] (r-\alpha  c t)+\sigma ^4 \left[\beta ^2 \left(2 n_r{}^2-1\right)+1\right]\,,\label{eq:ap-WEC-effective-proof2}
\end{equation}
implying that the WEC is equivalent to $S\geq0$.

\begin{figure}[h]
\includegraphics[width=\linewidth]{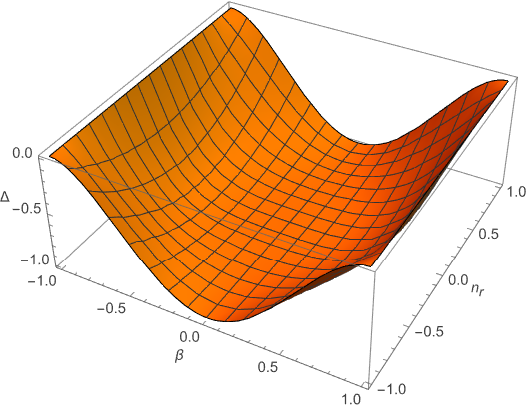}
 \caption{The discriminant $\Delta$ is plotted as a function of $\beta$ and $n_r$.}
 \label{fig:ap-discriminant}
\end{figure}

In order to study the WEC it is convenient to rewrite $S$ in terms of the variable $\gamma = r(r-\alpha c t)$, giving 
\begin{equation}
    S(\gamma)=8 (\beta  n_r-1)^2 \gamma^2+4  \sigma ^2 \left[2 n_r \beta\left(\beta  n_r-1\right)-\beta^2 +1\right] \gamma+\sigma ^4 \left[\beta ^2 \left(2 n_r{}^2-1\right)+1\right]\,,
\end{equation}
which is the equation of a parabola in $\gamma$.
The coefficient $8 (\beta  n_r-1)^2$ of the quadratic term is non-negative, since it is the square of a real quantity. This is therefore a concave-up parabola, and the overall sign of $S$ is determined by the discriminant $\Delta$ of the second-order algebraic equation $S(\gamma)=0$, which is given by
\begin{equation}
    \Delta = \sigma ^4 \left[\beta ^4-2 n_r^2 \left(\beta ^2-1\right) \beta ^2-1\right]\,.\label{eq:ap-discriminant}
\end{equation}
For the special case in which the quadratic term coefficient is zero, i.e. when $\beta n_r=1$, we have $S(\gamma)=4 \sigma^4$, so the WEC is satisfied.
As shown in Fig.~(\ref{fig:ap-discriminant}) the discriminant is never positive, which implies $S\geq0$ for any value of $\beta$ and $n_r$, i.e. for any timelike observer.

\bibliography{references}
\end{document}